\documentclass[conference]{IEEEtran}
\IEEEoverridecommandlockouts

\usepackage{cite}
\usepackage{amsmath,amssymb,amsfonts}
\usepackage{algorithmic}
\usepackage{graphicx}
\usepackage{textcomp}
\usepackage{xcolor}
\usepackage{tabu}
\usepackage{flushend}
\usepackage{algorithm}

\def\BibTeX{{\rm B\kern-.05em{\sc i\kern-.025em b}\kern-.08em
    T\kern-.1667em\lower.7ex\hbox{E}\kern-.125emX}}
\begin{document}

\title{Towards the Synthesis of Non-speech Vocalizations\\
}

\author{\IEEEauthorblockN{1\textsuperscript{st} Enjamamul Hoq}
\IEEEauthorblockA{\textit{University at Buffalo}\\
Buffalo, USA \\
ehoq@buffalo.edu}

\and
\IEEEauthorblockN{2\textsuperscript{th} Ifeoma Nwogu}
\IEEEauthorblockA{\textit{University at Buffalo}\\
Buffalo, USA \\
inwogu@buffalo.edu}

}

\maketitle

\begin{abstract}
In this report, we focus on the unconditional generation of infant cry sounds using the DiffWave framework, which has shown great promise in generating high-quality audio from noise. We use two distinct datasets of infant cries: the Baby Chillanto and the deBarbaro cry dataset. These datasets are used to train the DiffWave model to generate new cry sounds that maintain high fidelity and diversity. The focus here is on DiffWave's capability to handle the unconditional generation task.
\end{abstract}



\section{Dataset}
The DiffWave model is trained using two different infant cry datasets:
\begin{itemize}
    \item Baby Chillanto Dataset: This dataset\cite{reyes2008validation,rosales2015classifying} contains 2274 samples of infant cries, each approximately 1 second long and resampled at 16kHz during training. This dataset has five different categories of infant cry signals. They are as follows: deaf, hearing, asphyxia, hungry and pain.
    \item deBarbaro Cry Dataset: This dataset\cite{yao2022infant} is significantly larger, with 44129 samples with a cry and no cry labels. We only used 22232 cry samples each 5 seconds long. We resampled the cry signals from 22KHz to 16KHz during training.
\end{itemize}

\section{Methodology}

The DiffWave model is built on a diffusion probabilistic framework, where the key idea is to transform data gradually by adding noise through a \textit{diffusin process} also known as forward process and then reverse this transformation using a \textit{reverse process} to generate new data. This process allows DiffWave to model the structure of audio waveforms in a generative way. \textbf{Figure 1} shows the forward and reverse process of the DiffWave model.

\begin{figure}[htbp]
    \centering
    \includegraphics[width=\linewidth]{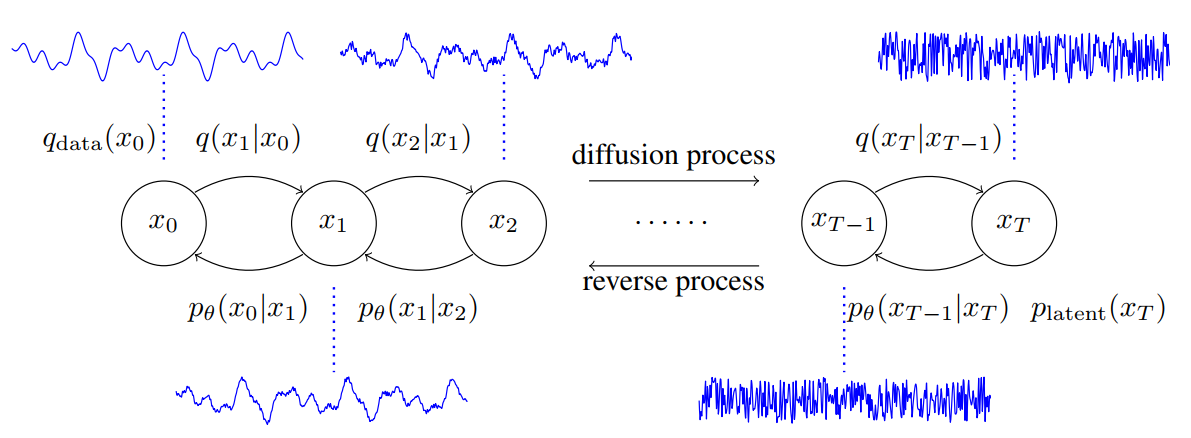}
    \caption{The diffusion and reverse process in diffusion probabilistic models. Figure taken from \cite{kong2020diffwave}.}
    \label{fig:fig2}
\end{figure}

\subsection{Forward Process}

In the forward diffusion process, noise is gradually added to the data through a Markov chain and transforms the clean data \( x_0 \) into a latent variable \( x_T \), which is close to random Gaussian noise. At each step \( t \), small Gaussian noise is added to the previous step’s data. This process is modeled as follows:

\begin{equation}
q(x_1, \dots, x_T | x_0) = \prod_{t=1}^{T} q(x_t | x_{t-1})
\end{equation}

where the transition probability \( q(x_t | x_{t-1}) \) is defined as:

\begin{equation}
q(x_t | x_{t-1}) = \mathcal{N}(x_t; \sqrt{1-\beta_t} x_{t-1}, \beta_t I)
\end{equation}

In this formulation:
\begin{itemize}
    \item \( \beta_t \) is a small positive constant that controls the amount of noise added at each step, known as the \textit{variance schedule}.
    \item \( \sqrt{1 - \beta_t} \) is a scaling factor applied to the data from the previous step \( x_{t-1} \).
    \item \( \beta_t I \) represents the variance of the Gaussian noise added at each step.
\end{itemize}

The role of this forward process is to add small Gaussian noise to the data step by step, gradually transforming the original data \( x_0 \) into a noise-like latent representation \( x_T \). By the time we reach \( T \) steps, the data is close to an isotropic Gaussian distribution \( \mathcal{N}(0, I) \).

\subsection{Reverse Process}

The reverse process aims to recover the original data \( x_0 \) from the latent variable \( x_T \), which has been transformed into noise. This reverse process is learned by the model during training and is modeled as another Markov chain, starting from \( x_T \) and iteratively generating the data back to \( x_0 \).

The reverse process is expressed as:

\begin{equation}
p_\theta(x_0, \dots, x_{T-1} | x_T) = \prod_{t=1}^{T} p_\theta(x_{t-1} | x_t)
\end{equation}

where the conditional distribution at each reverse step \( p_\theta(x_{t-1} | x_t) \) is parameterized as:

\begin{equation}
p_\theta(x_{t-1} | x_t) = \mathcal{N}(x_{t-1}; \mu_\theta(x_t, t), \sigma_\theta(x_t, t)^2 I)
\end{equation}

In this reverse process:
\begin{itemize}
    \item \( \mu_\theta(x_t, t) \) is the predicted mean of the distribution that the model learns to denoise \( x_t \), based on the current noisy data and the diffusion step \( t \).
    \item \( \sigma_\theta(x_t, t) \) is the predicted variance.
\end{itemize}

At each step, the model removes the Gaussian noise that was added in the forward process, producing cleaner data. This reverse process continues until the model has denoised the data from \( x_T \) pure gaussian noise back to \( x_0 \), the original data. The goal of this reverse process is to remove the noise added during the diffusion process and reconstruct the original, structured audio.

\subsection{The Loss Function}

 The training objective is derived from the \textit{Evidence Lower Bound} (ELBO) and simplifies to a Mean Squared Error (MSE) loss between the true noise \( \epsilon \) and the noise predicted by the model \( \epsilon_\theta(x_t, t) \). This loss function ensures that the model learns to accurately reverse the diffusion process by correctly predicting the noise at each step. Detail derivation is provided in the \textbf{appendix A} section of \cite{kong2020diffwave}. 

The loss function is given by:

\begin{equation}
L(\theta) = \mathbb{E}_{x_0, t, \epsilon} \left[ \| \epsilon - \epsilon_\theta(\sqrt{\bar{\alpha}_t} x_0 + \sqrt{1 - \bar{\alpha}_t} \epsilon, t) \|_2^2 \right]
\end{equation}

where:
\begin{itemize}
    \item \( \epsilon \) is the Gaussian noise added during the forward diffusion process,
    \item \( \epsilon_\theta(x_t, t) \) is the noise predicted by the model at step \( t \),
    \item \( \alpha_t \) is a function of the variance schedule \( \beta_t \), specifically \( \alpha_t = 1 - \beta_t \) and \( \bar{\alpha}_t \) is the cumulative product of \( \alpha_t \) across all time steps.
\end{itemize}

\section{Model Architecture}

DiffWave’s architecture is a non-autoregressive, feed-forward convolutional network that draws inspiration from WaveNet but introduces key differences. Unlike WaveNet, which generates audio sequentially in an autoregressive manner, DiffWave generates the entire audio sequence in parallel, making it significantly faster and more efficient for long-form audio tasks. \textbf{Figure 2} shows the DiffWave framework.

\begin{figure}[htbp]
    \centering
    \includegraphics[width=.6\linewidth]{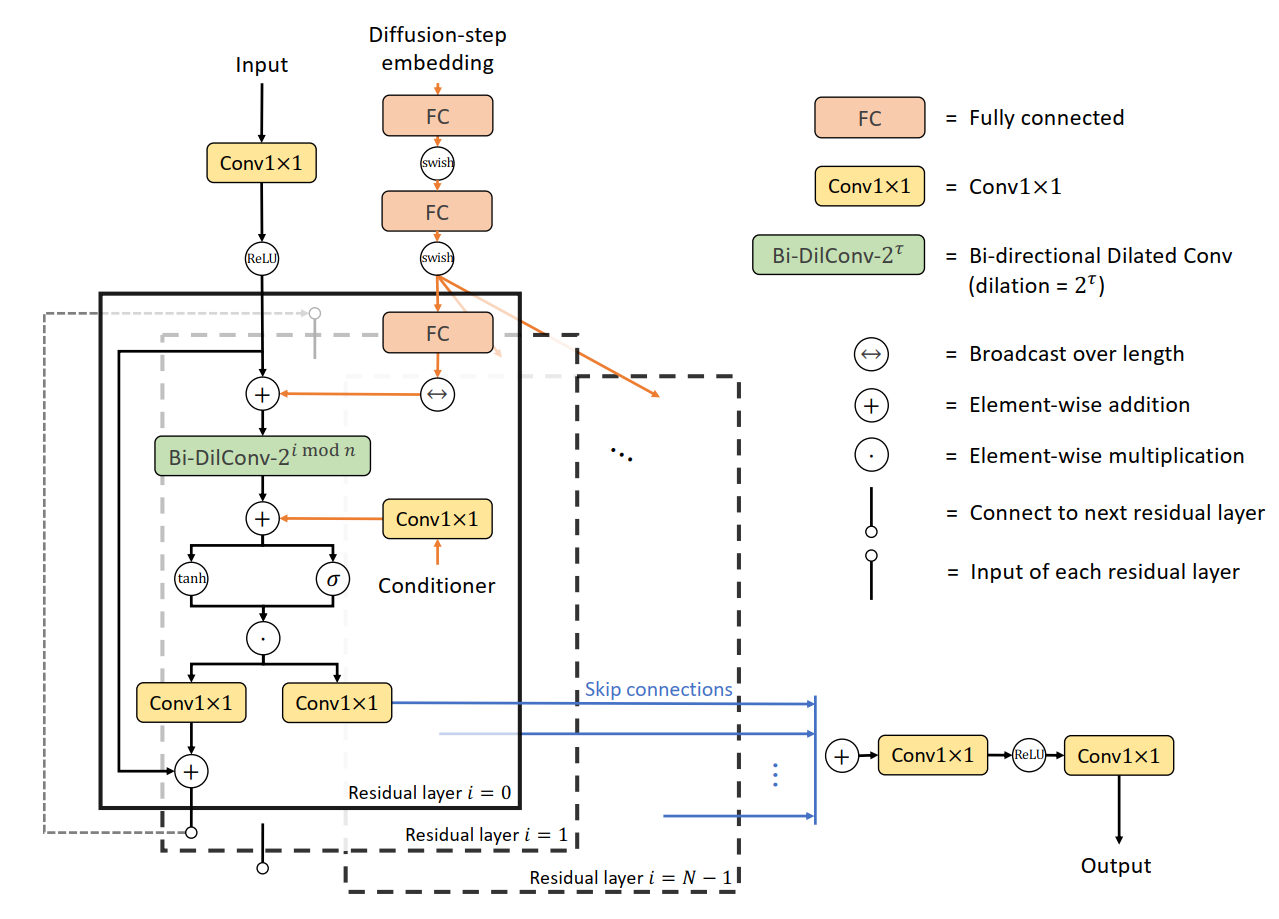}
    \caption{The network architecture of DiffWave in modeling $\epsilon_\theta$. Figure taken from \cite{kong2020diffwave}.}
    \label{fig:fig1}
\end{figure}

\subsection{Dilated Convolutions}
Both models use dilated convolutions to increase the receptive field exponentially, capturing both local and long-range dependencies. However, while WaveNet uses causal convolutions to maintain temporal order, DiffWave uses bidirectional dilated convolutions, enabling it to consider both past and future context and generate audio in parallel.

\subsection{Residual Layers}
Similar to WaveNet, DiffWave incorporates residual layers with dilated convolutions and nonlinear activations like ReLU and swish. Skip connections ensure stable training and efficient gradient flow, but DiffWave applies these in a non-autoregressive setting.

\subsection{Conditioning on Diffusion Steps}

DiffWave conditions its denoising process on the diffusion step \( t \) by using a sinusoidal position embedding, similar to the one used in transformer architectures. This embedding is a 128-dimensional vector that encodes the diffusion step using both sine and cosine functions over different frequencies. The embedding equation is given as:


\begin{equation}
\begin{aligned}
    t_{\text{embedding}} = \Bigg[ &\sin\left(10^{\frac{0 \times 4}{63}} t\right), \cdots, \sin\left(10^{\frac{63 \times 4}{63}} t\right), \\
    &\cos\left(10^{\frac{0 \times 4}{63}} t\right), \cdots, \cos\left(10^{\frac{63 \times 4}{63}} t\right) \Bigg]
\end{aligned}
\end{equation}

This embedding vector is then passed through fully connected (FC) layers before being added to the input of each residual layer. This allows each residual layer to be aware of the current diffusion step and adjust its behavior based on how much noise remains in the signal at that stage.

The reverse process at each step \( p_\theta(x_{t-1} | x_t) \) is conditioned on both the noisy input \( x_t \) and the diffusion step \( t \), enabling the model to remove noise dynamically. The sinusoidal embedding helps the network learn complex patterns over time and allows for the generation of audio sequences that improve progressively with each reverse diffusion step.

\section{Training Setup}
The training setup for unconditional generation in DiffWave focuses on transforming random noise into structured audio without any conditional input. The model was trained on the two infant cry datasets. DiffWave uses 200 diffusion steps, and the network consists of 30 residual layers with bidirectional dilated convolutions to capture both local and global dependencies. The variance schedule \( \beta_t \) was linearly spaced between \( [1 \times 10^{-4}, 0.02] \), with adjustments for smaller model configurations.

The model was optimized using the Adam optimizer with a fixed learning rate of \( 2 \times 10^{-4} \) and a batch size of 16. The training was performed for 500K steps using a single NVIDIA 3090 RTX 24GB GPU for each dataset. It takes an average of $3$ days to complete the training. After 150k steps, the model starts to generate clear cry sounds for the deBarbaro dataset. Fast sampling techniques, reducing reverse steps from 200 to as few as 6, were also employed to speed up generation without compromising audio quality.

\section{Training Process}

The training process is based on maximizing the variational lower bound (ELBO) to ensure the model learns how to reverse the noise addition process. The likelihood \( p_\theta(x_0) \) is intractable in general, so it is approximated by maximizing the ELBO:


\begin{equation}
\begin{aligned}
    \mathbb{E}_{q_{\text{data}}(x_0)} \log p_\theta(x_0) 
    &\geq \mathbb{E}_{q(x_0, \dots, x_T)} \Bigg[ \log \frac{p_\theta(x_0, \dots, x_{T-1} | x_T)}{q(x_1, \dots, x_T | x_0)} \\
    &\quad + \log p_{\text{latent}}(x_T) \Bigg]
\end{aligned}
\end{equation}

The ELBO is derived to minimize the gap between the true posterior and the approximate posterior. Ho et al. \cite{ho2020denoising} parameterize this process in a way that allows it to be computed in closed form, avoiding Monte Carlo estimates and accelerating training. The parameterization defines constants based on the variance schedule \( \{\beta_t\}_{t=1}^{T} \):

\begin{equation}
    \alpha_t = 1 - \beta_t, \quad \bar{\alpha}_t = \prod_{s=1}^{t} \alpha_s, \quad \tilde{\beta}_t = \frac{1 - \bar{\alpha}_{t-1}}{1 - \bar{\alpha}_t} \beta_t
\end{equation}

The model learns the parameterized means and variances:

\begin{equation}
    \mu_\theta(x_t, t) = \frac{1}{\sqrt{\alpha_t}} \left( x_t - \frac{\beta_t}{\sqrt{1 - \bar{\alpha}_t}} \epsilon_\theta(x_t, t) \right), \quad \sigma_\theta(x_t, t) = \tilde{\beta}_t
\end{equation}

where \( \epsilon_\theta(x_t, t) \) is a neural network predicting the noise.

\begin{algorithm}[H]
\caption{Training algorithm from \cite{kong2020diffwave}}
\begin{algorithmic}[1]
    \FOR{$i = 1$ to $N_{\text{iter}}$}
        \STATE Sample $x_0 \sim q_{\text{data}}, \epsilon \sim \mathcal{N}(0, I)$
        \STATE $t \sim \operatorname{Uniform}(\{1, \cdots, T\})$
        \STATE Take gradient step on
        \STATE $\nabla_\theta\left\|\epsilon - \epsilon_\theta\left(\sqrt{\bar{\alpha}_t} x_0 + \sqrt{1-\bar{\alpha}_t} \epsilon, t\right)\right\|_2^2$
        \STATE according to Eq.~(5)
    \ENDFOR
\end{algorithmic}
\end{algorithm}

\section{Sampling Process}

The sampling process is the reverse of the diffusion process. The model starts with pure noise and progressively denoises it to generate data. The generative process is defined as follows:

\begin{equation}
    p_\theta(x_0, \dots, x_{T-1} | x_T) = \prod_{t=1}^{T} p_\theta(x_{t-1} | x_t)
\end{equation}

The reverse process starts by sampling \( x_T \sim \mathcal{N}(0, I) \) and iterating over the following steps for each \( t \) from \( T \) to 1:

\begin{equation}
    x_{t-1} \sim p_\theta(x_{t-1} | x_t) = \mathcal{N}(x_{t-1}; \mu_\theta(x_t, t), \sigma_\theta(x_t, t)^2 I)
\end{equation}

where \( \mu_\theta(x_t, t) \) and \( \sigma_\theta(x_t, t) \) are computed using the learned parameters. At each step, the model progressively denoises the sample, generating structured data from Gaussian noise. The output at \( t=0 \), denoted \( x_0 \), is the final generated data sample.

\begin{algorithm}
\caption{Sampling algorithm from \cite{kong2020diffwave}}
\begin{algorithmic}[1]
    \STATE Sample $x_T \sim p_{\text{latent}} = \mathcal{N}(0, I)$
    \FOR{$t = T$ to $1$}
        \STATE Compute $\mu_\theta\left(x_t, t\right)$ and $\sigma_\theta\left(x_t, t\right)$ using Eq.~(9)
        \STATE Sample $x_{t-1} \sim p_\theta\left(x_{t-1} \mid x_t\right)$
        \STATE $\mathcal{N}\left(x_{t-1} ; \mu_\theta\left(x_t, t\right), \sigma_\theta\left(x_t, t\right)^2 I\right)$
    \ENDFOR
    \STATE return $x_0$
\end{algorithmic}
\end{algorithm}

\section{Results}
Figure 3 shows the Mel-spectrogram of the unconditionally generated infant cry signals using the DiffWave model. Generated audio samples have 16KHz sampling rate. 80-channel Mel-spectrogram is computed on
25-millisecond windows with a stride of 10 milliseconds.  


\begin{figure}[htbp]
\centerline{\includegraphics[width=\linewidth]{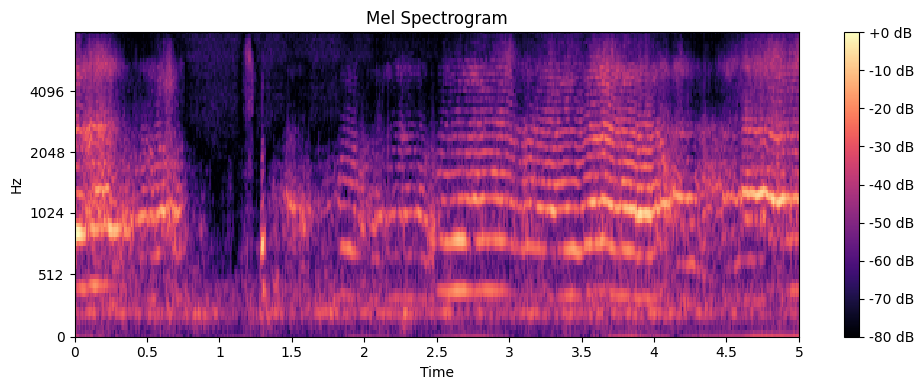}}
\centerline{\includegraphics[width=\linewidth]{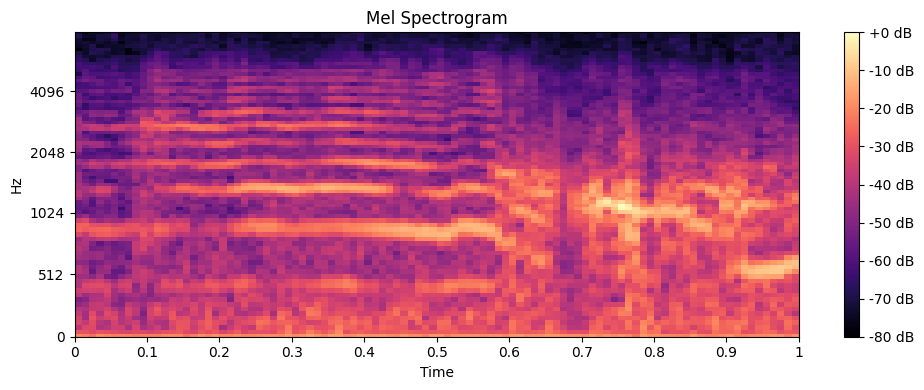}}
\caption{Generated samples from model trained on a) deBarbaro dataset b) Baby Chillanto dataset.}
\label{fig:mel}
\end{figure}

\newpage
\bibliographystyle{IEEEtran}
\bibliography{main.bib}

\begin{thebibliography}{1}
\providecommand{\url}[1]{#1}
\csname url@samestyle\endcsname
\providecommand{\newblock}{\relax}
\providecommand{\bibinfo}[2]{#2}
\providecommand{\BIBentrySTDinterwordspacing}{\spaceskip=0pt\relax}
\providecommand{\BIBentryALTinterwordstretchfactor}{4}
\providecommand{\BIBentryALTinterwordspacing}{\spaceskip=\fontdimen2\font plus
\BIBentryALTinterwordstretchfactor\fontdimen3\font minus
  \fontdimen4\font\relax}
\providecommand{\BIBforeignlanguage}[2]{{%
\expandafter\ifx\csname l@#1\endcsname\relax
\typeout{** WARNING: IEEEtran.bst: No hyphenation pattern has been}%
\typeout{** loaded for the language `#1'. Using the pattern for}%
\typeout{** the default language instead.}%
\else
\language=\csname l@#1\endcsname
\fi
#2}}
\providecommand{\BIBdecl}{\relax}
\BIBdecl

\bibitem{reyes2008validation}
O.~F. Reyes-Galaviz, S.~D. Cano-Ortiz, and C.~A. Reyes-Garc{\'\i}a,
  ``Validation of the cry unit as primary element for cry analysis using an
  evolutionary-neural approach,'' in \emph{2008 Mexican International
  Conference on Computer Science}.\hskip 1em plus 0.5em minus 0.4em\relax IEEE,
  2008, pp. 261--267.

\bibitem{rosales2015classifying}
A.~Rosales-P{\'e}rez, C.~A. Reyes-Garc{\'\i}a, J.~A. Gonzalez, O.~F.
  Reyes-Galaviz, H.~J. Escalante, and S.~Orlandi, ``Classifying infant cry
  patterns by the genetic selection of a fuzzy model,'' \emph{Biomedical Signal
  Processing and Control}, vol.~17, pp. 38--46, 2015.

\bibitem{yao2022infant}
X.~Yao, M.~Micheletti, M.~Johnson, E.~Thomaz, and K.~de~Barbaro, ``Infant
  crying detection in real-world environments,'' in \emph{ICASSP 2022-2022 IEEE
  International Conference on Acoustics, Speech and Signal Processing
  (ICASSP)}.\hskip 1em plus 0.5em minus 0.4em\relax IEEE, 2022, pp. 131--135.

\bibitem{kong2020diffwave}
Z.~Kong, W.~Ping, J.~Huang, K.~Zhao, and B.~Catanzaro, ``Diffwave: A versatile
  diffusion model for audio synthesis,'' \emph{arXiv preprint
  arXiv:2009.09761}, 2020.

\bibitem{ho2020denoising}
J.~Ho, A.~Jain, and P.~Abbeel, ``Denoising diffusion probabilistic models,''
  \emph{Advances in neural information processing systems}, vol.~33, pp.
  6840--6851, 2020.

\end{thebibliography}

\end{document}